\documentclass[a4paper,nobibnotes,nofootinbib,superscriptaddress]{revtex4}

\usepackage{amssymb}
\usepackage{amsmath}
\usepackage{epsfig}

\newcommand{\be}{\begin{equation}}
\newcommand{\ee}{\end{equation}}
\newcommand{\bea}{\begin{eqnarray}}
\newcommand{\eea}{\end{eqnarray}}

\newcommand{\g}{\gamma}
\newcommand{\f}{\frac}

\newcommand{\intc}[1]{{\int\frac{d#1}{2i\pi}}}
\newcommand\lr[1]{{\left({#1}\right)}}

\begin{document}

\title{Gaps between jets in hadronic collisions}
\author{O. Kepka}\email{kepkao@fzu.cz}
\affiliation{IPNP, Faculty of Mathematics and Physics, Charles University, Prague, Czech Republic}
\affiliation{Center for Particle Physics, Institute of Physics, Academy of Science, Prague, Czech Republic}
\author{C. Marquet}\email{cyrille.marquet@cern.ch}
\affiliation{Physics Department, Theory Unit, CERN, 1211 Gen\`eve 23, Switzerland}
\author{C. Royon}\email{christophe.royon@cea.fr}
\affiliation{IRFU/Service de physique des particules, CEA/Saclay, 91191 Gif-sur-Yvette cedex, France}

\begin{abstract}

We propose a model to describe diffractive events in hadron-hadron collisions where a rapidity gap is surrounded by two jets. The hard color-singlet object exchanged in the t-channel and responsible for the rapidity gap is described by the pQCD Balitsky-Fadin-Kuraev-Lipatov Pomeron, including corrections due to next-to-leading logarithms. We allow the rapidity gap to be smaller than the inter-jet rapidity interval, and the corresponding soft radiation is modeled using the HERWIG Monte Carlo. Our model is able to reproduce all Tevatron data, and allows to estimate the jet-gap-jet cross section at the LHC.

\end{abstract}

\maketitle

\section{Introduction}

A large effort has been devoted to understand the QCD dynamics of rapidity gaps in jet events since such processes were observed in $p+\bar{p}$ collisions at the Tevatron more than 10 years ago \cite{d0,cdf}. While describing diffractive processes in QCD has been a challenge for many years, the presence of a hard scale in so-called {\it jet-gap-jet} events for instance, brings hope that one could be able to understand these with perturbative methods. However, after many theoretical investigations, there is still no consensus on what the relevant QCD mechanism really is.

In a hadron-hadron collision, a jet-gap-jet event features a large rapidity gap with a high-$p_T$ jet on each side ($p_T\!\gg\!\Lambda_{QCD}$). Across the gap, the object exchanged in the $t-$channel is color singlet and carries a large momentum transfer, and when the rapidity gap is sufficiently large the natural candidate in perturbative QCD is the Balitsky-Fadin-Kuraev-Lipatov (BFKL) Pomeron \cite{bfkl}. Of course the collision energy $\sqrt{s}$ should be big ($\sqrt{s}\gg E_T$) in order for jets to be produced along with a large rapidity gap. Such events are expected to be produced copiously in $p+p$ collisions at the LHC.

To compute the jet-gap-jet process in the BFKL framework, one has first to address the problem of coupling the BFKL Pomeron to partons, as opposed to colorless particles. Indeed, BFKL calculations usually use the fact that impact factors, which describe the coupling of incoming and outgoing particles to the BFKL Pomeron, vanish when attached to gluons with no transverse momentum. This is a property of colorless impact factors. For instance, this is what allows to turn the Feynman-diagram calculation of the BFKL Pomeron into a conformal-invariant Green function \cite{lipatov}. Consequently, this BFKL Green function cannot be hooked to colored particles, and should be modified accordingly first. The Mueller-Tang (MT) prescription \cite{muellertang} is widely used in the literature to couple the BFKL Pomeron to quarks and gluons.

On the phenomenological side, the original parton-level MT calculation was not sufficient to describe the Tevatron data. A first attempt to improve it was proposed in \cite{cfl}, parton showering and hadronization were taken into account using the HERWIG Monte Carlo program \cite{herwig}. An agreement with data could only be obtained if the leading-logarithmic (LL) BFKL calculation was done with a fixed value of the coupling constant $\alpha_S,$ which is not satisfactory, as next-to-leading logarithmic (NLL) BFKL corrections are known to be important. In addition, only the leading conformal spin ($p=0$) was taken into account. In
\cite{rikard}, it was shown that a good description of the data could be obtained when some NLL corrections were numerically taken into account in an effective way \cite{fakenll}, but the full NLL-BFKL kernel
\cite{nllbfkl} could still not be implemented. As a result, these tests on the relevance of the BFKL dynamics were not conclusive.

In the most recent phenomenological work on the subject \cite{us}, the full NLL-BFKL kernel was implemented including all conformal spins, along with the collinear improvements necessary to remove spurious singularities and obtain meaningful results \cite{salam,ccs}. However, the results of \cite{us} remained at parton level, therefore the fact that the rapidity interval between the jets can be larger than the rapidity gap could not be implemented. The D0 measurement could nevertheless be reasonably well described, while the CDF data was not considered. The purpose of this letter is to improve the model by taking into account parton showering, hadronization effects and jet reconstruction, which is necessary to make more precise comparisons with data. We will interface the collinearly-improved NLL-BFKL parton-level results of \cite{us} with the HERWIG Monte Carlo, as was done in \cite{cfl} with the LL-BFKL calculation. We shall compare our results to both D0 and CDF data.

The plan of the letter is as follows. In section II, we recall the phenomenological NLL-BFKL formulation of the
jet-gap-jet cross section and in section III, we explain how it is embedded into the HERWIG Monte Carlo program. In section IV, we present successful comparisons with all Tevatron data, which allow to fix the absolute normalization in our model. Predictions for the jet-gap-jet cross section at the LHC are presented in Section V. Section VI is devoted to conclusions and outlook.

\section{The jet-gap-jet cross section in the BFKL framework}

\begin{figure}[t]
\begin{center}
\epsfig{file=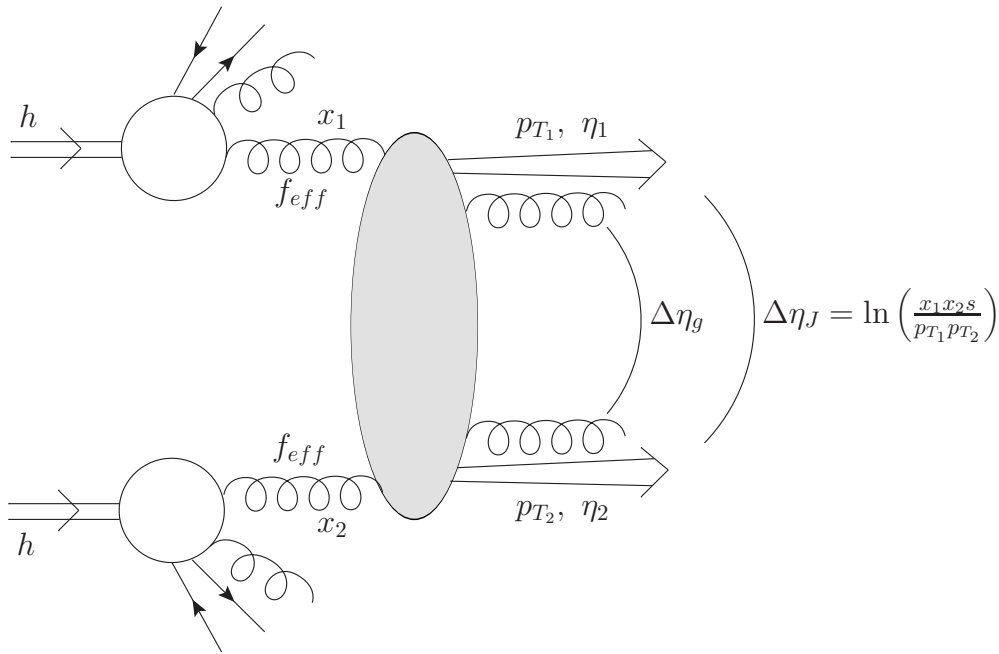,width=13cm}
\caption{Production of two jets surrounding a large rapidity gap in a hadron-hadron collision. $\sqrt{s}$ denotes the collision energy, $p_{T_1}$ ($\eta_1$) and $p_{T_2}$ ($\eta_2$) the transverse momenta (rapidities) of the jets and $x_1$ and $x_2$ are their longitudinal momentum fraction with respect to the incident hadrons. The rapidity interval between the jets $\Delta\eta_J$ is bigger than the rapidity gap
$\Delta\eta_g$.}
\label{jetgapjet}
\end{center}
\end{figure}

The production of a rapidity gap between two outgoing jets in a hadron-hadron collision is pictured in
Fig.~\ref{jetgapjet}, with the different kinematic variables. We denote $\sqrt{s}$ the collision energy,
$p_{T_1}$ and $p_{T_2}$ the transverse momenta of the two jets and $x_1$ and $x_2$ their longitudinal fraction of momentum with respect to the incident hadrons. The rapidity interval between the two jets is $\Delta\eta_J=\ln(x_1x_2s/p_{T_1}p_{T_2})$. At the parton level (see Fig.1 in \cite{us}),
$p_{T_1}=-p_{T_2}=p_T$, and the rapidity gap coincides with the rapidity interval
$\Delta\eta\!=\!\ln(x_1x_2s/p_T^2)$ between the outgoing partons that will initiate the jets. The hadronization of the partons into jets reduces the size of the rapidity gap to $\Delta\eta_g$.

In this section, we deal with the parton-level cross section
\begin{equation}
\frac{d\sigma^{pp\to XJJY}}{dx_1 dx_2 dp_T^2} = {\cal S}f_{eff}(x_1,p_T^2)f_{eff}(x_2,p_T^2)
\frac{d\sigma^{gg\rightarrow gg}}{dp_T^2},
\label{jgj}
\end{equation}
where the functions $f_{eff}(x,p_T^2)$ are effective parton distributions that resum the leading logarithms
$\log(p_T^2/\Lambda_{QCD}^2)$. They have the form
\be
f_{eff}(x,\mu^2)=g(x,\mu^2)+\f{C_F^2}{N_c^2}\lr{q(x,\mu^2)+\bar{q}(x,\mu^2)}\ ,
\label{pdfs}
\ee
where $g$ (respectively $q$, $\bar{q}$) is the gluon (respectively quark, antiquark) distribution function in the incoming hadrons, and evolves according to DGLAP evolution \cite{dglap}. Even though the process we consider involves moderate values of $x_1$ and $x_2$ and the perturbative scale $p_T^2\gg\Lambda_{QCD}^2,$ which we have chosen as the factorization scale, the cross section \eqref{jgj} does not obey collinear factorization. This is due to possible secondary soft interactions between the colliding hadrons which can fill the rapidity gap. Therefore, in \eqref{jgj}, the collinear factorization of the parton distributions $f_{eff}$ is corrected with the so-called gap-survival probability ${\cal S}$, which we assume depends only on $\sqrt{s}$ as in standard diffractive calculations. Since the soft interactions happen on much longer time scales, the factor ${\cal S}$ is factorized from the hard part
$d\sigma^{gg\rightarrow gg}/dp_T^2$. This hard cross section is given by
\begin{equation}
\frac{d \sigma^{gg\rightarrow gg}}{dp_T^2}=\frac{1}{16\pi}\left|A(\Delta\eta,p_T^2)\right|^2
\label{hardpart}
\end{equation}
in terms of the $gg\to gg$ scattering amplitude $A(\Delta\eta,p_T^2).$ The two measured jets are initiated by the final-state gluons (or quarks), parton showering and hadronization effects will be discussed in the next section.

In the following, we consider the high-energy limit in which the rapidity gap $\Delta\eta$ is assumed to be very large. The BFKL framework allows to compute the $gg\to gg$ amplitude in this regime, and the result is known up to NLL accuracy. We note that there exist other QCD-based approaches to compute the jet-gap-jet cross section \cite{sll}. Let us first point out that in general collinear and $k_T$-factorization are two distinct schemes to factorize a hard process from a soft process (as is the case for the proton structure function
$F_2$), and should not be mixed. But the process we are investigating is different: collinear factorization is used to separate the hard part from the soft part, and $k_T$-factorization is only used within the hard part itself. It allows to factorize the amplitude $A(\Delta\eta,p_T^2)$ into three hard pieces: two impact factors defined order-by-order with respect to $\alpha_S,$ and the BFKL Green function where a resummation of leading (and next-leading) logarithms is performed.

Since in our calculation the BFKL Pomeron is coupled to quarks or gluons, the BFKL Green function cannot be used as it is and should be modified. The transformation proposed in \cite{muellertang} is based on the fact that one should recover the analiticity of the Feynman diagrams. It was later argued that this prescripion corresponds to a deformed representation of the BFKL kernel that indeed could be coupled to colored particles and for which the bootstrap relation is fullfiled \cite{barlip}. Applying the MT prescription at NLL leads to
\begin{equation}
A(\Delta\eta,p_T^2)=\frac{16N_c\pi\alpha_S^2(p_T^2)}{C_Fp_T^2}\sum_{p=-\infty}^\infty\intc{\g}
\frac{[p^2-(\g-1/2)^2]\exp\left\{\bar\alpha(p_T^2)\chi_{eff}[2p,\g,\bar\alpha(p_T^2)] \Delta \eta\right\}}
{[(\g-1/2)^2-(p-1/2)^2][(\g-1/2)^2-(p+1/2)^2]} 
\label{jgjnll}
\end{equation}
with the complex integral running along the imaginary axis from $1/2\!-\!i\infty$ 
to $1/2\!+\!i\infty,$ and with only even conformal spins contributing to the sum \cite{leszek}.
The running coupling is given by
\be
\bar\alpha(p_T^2)=\f{\alpha_S(p_T^2)N_c}{\pi}=
\left[b\log\lr{p_T^2/\Lambda_{QCD}^2}\right]^{-1}\ ,\quad b=\f{11N_c-2N_f}{12N_c}\ .
\ee
It is important to note that in formula \eqref{jgjnll}, we used the leading-order non-forward quark and gluon impact factors. We point out that the next-to-leading-order impact factors are known \cite{ifnlo}, and that in principle a full NLL analysis is feasible, but this goes beyond the scope of our study.

The NLL-BFKL effects are phenomenologically taken into account by the effective kernels
$\chi_{eff}(p,\g,\bar\alpha).$ For $p=0,$ the scheme-dependent NLL-BFKL kernels provided by the regularisation procedure $\chi_{NLL}\lr{\g,\omega}$ depend on $\omega,$ the Mellin variable conjugate to $\exp(\Delta\eta).$ In each case, the NLL kernels obey a {\it consistency condition} \cite{salam} which allows to reformulate the problem in terms of $\chi_{eff}(\g,\bar\alpha)$ (see also \cite{ccs,singnll} for different approaches). The effective kernel $\chi_{eff}(\g,\bar\alpha)$ is obtained from the NLL kernel $\chi_{NLL}\lr{\g,\omega}$ by solving the implicit equation $\chi_{eff}=\chi_{NLL}\lr{\g,\bar\alpha\ \chi_{eff}}$. In
\cite{nllmnjus,nllmnjthem}, the regularisation procedure has been extended to non-zero conformal spins and the kernel $\chi_{NLL}\lr{p,\g,\omega}$ was obtained from the results of \cite{kotlip}. The formulae needed to compute it can be found in the appendix of \cite{nllmnjus} (in the present study we shall use the S4 scheme in which $\chi_{NLL}$ is supplemented by an explicit $\bar\alpha$ dependence, the results in the case of the S3 scheme are similar). Then the effective kernels $\chi_{eff}(p,\g,\bar\alpha)$ are obtained from the NLL kernel by solving the implicit equation:
\be
\chi_{eff}=\chi_{NLL}\lr{p,\g,\bar\alpha\ \chi_{eff}}\ .
\label{eff}
\ee

Similar NLL-BFKL phenomenological studies have been carried out with Mueller-Navelet jets in hadron-hadron collisions \cite{nllmnjus,nllmnjthem}, forward jet production in deep inelastic scattering
\cite{nllfjus,nllfjthem}, and the proton structure function \cite{nllf2}. While in the $F_2$ analysis the NLL corrections did not really improve the BFKL description, it was definitively the case in the forward-jet study. In the Mueller-Navelet jet case, NLL corrections dramatically change the predictions, even more so in the full calculation when NLO impact factors are also implemented \cite{nllmnjfull}. In fact, these results cast strong doubts on the fact that Mueller-Navelet jets are a good observable to unambiguously observe BFKL effects, leaving the jet-gap-jet measurement as perhaps the new candidate.

In the LL-BFKL case that we consider for comparisons, the formula for the jet-gap-jet cross section is formally the same as the NLL one, with the following substitutions in \eqref{jgjnll}:
\be
\chi_{eff}(p,\g,\bar\alpha)\rightarrow\chi_{LL}(p,\g)
=2\psi(1)-\psi\lr{1-\g+\f{|p|}2}-\psi\lr{\g+\f{|p|}2}\ ,
\hspace{1cm}\bar\alpha(k^2)\rightarrow\bar\alpha=\mbox{const. parameter} ,
\label{chill}\ee
where $\psi(\g)\!=\!d\log\Gamma(\g)/d\g$ is the logarithmic derivative of the Gamma function. In this case, the coupling $\bar\alpha$ is a priori a parameter. We choose to fix it to the value 0.16 obtained in
\cite{nllfjus} by fitting the forward jet data from HERA. This unphysically small value of the coupling is indicative of the slower Bjorken-$x$ dependence of the forward-jet data compared to the LL-BFKL cross section, when used with a reasonable $\bar\alpha$ value. And in fact, the value $\bar\alpha=0.16$ mimics the slower energy dependence of NLL-BFKL cross section (in this case the average value of $\bar\alpha$ is about 0.25), which in the forward-jet case is consistent with data. Therefore in both the LL- and NLL-BFKL cases, one deals with one-parameter formulae: the absolute normalization which is not under control. In the NLL case, this is due to the fact the we do not use NLO impact factors.

Finally, to compute the cross section \eqref{jgj}, we use CTEQ parton distribution functions \cite{cteq}, 
and we take $S=0.1$ for the gap-survival probability at the Tevatron and $S=0.03$ at the LHC. More details on the parton-level computations can be found in \cite{us}, such as the importance of the different conformal spins in \eqref{jgjnll}, or the uncertainty due to the choice of the renormalization scale. In this work, the goal is to obtain hadron-level results by interfacing \eqref{jgj} with the HERWIG event generator.

\section{Implementation of the NLL-BFKL formula in HERWIG}

The parton-level calculation presented in the previous section leads by definition to a gap size $\Delta\eta$ equal to the interval in rapidity between the partons that initiate the jets. At particle level, it is no longer true. Due to QCD radiation and hadronisation, the jets have a finite size, and the gap size
$\Delta\eta_g$ is smaller than the difference in rapidity between the two jets $\Delta\eta_J$ (see
Fig.~\ref{jetgapjet}). This has an important consequence: to be able to compare the NLL-BFKL jet-gap-jet cross sections with CDF and D0 measurements at the Tevatron, it is needed to embed our formulae in a Monte Carlo code. For instance, the D0 collaboration selects events with a gap devoid of any activity in the $[-1,1]$ region in rapidity while they require the jets to be separated by at least 4 units in rapidity. To take into account these effects, we implemented the NLL-BFKL cross section in the HERWIG Monte Carlo, and made our analysis at the particle level, after hadronisation. Since this procedure of going from parton-level to hadron-level is quite sensitive to the way jets are reconstructed, we use the same jet algorithm as experimentally used.

Practically, in order to implement our formalism in HERWIG, we modified the HWHSNM function which implements the matrix element squared for color-singlet parton-parton scattering \cite{herwig}. Formula
\eqref{hardpart}, which gives the BFKL $d\sigma/dp_T^2$ cross section is too complicated to be implemented
directly in HERWIG since it involves an integration in the complex plane over $\gamma$, and it would
take too much computing time to generate many events. To avoid this issue, we parameterized $d\sigma/dp_T^2$ as a function of the parton $p_T$ and $\Delta\eta$ between both partons at generator level. Denoting 
$z(p_T^2)=\bar\alpha(p_T^2)\Delta\eta/2$, the parametrization used is
\be
\frac{d \sigma}{dp_T^2}=\frac{\alpha_S^4(p_T^2)}{4\pi p_T^4}  \left[ a + b p_T + c \sqrt{p_T}
+ (d + e p_T + f \sqrt{p_T})\times z + (g + h p_T)\times z^2 +
(i + j \sqrt{p_T})\times z^3 + \exp(k + l z) \right]\ .
\label{formulafit}
\ee

This formula is purely phenomenological, not motivated by theory, and was just introduced to obtain
a very good $\chi^2$ while fitting \eqref{formulafit} to the full expression of $d\sigma/dp_T^2$. To perform the fit, 2330 points were used for parton $p_T$ ranging from 10 to 120 GeV, and $\Delta\eta$ up to 10. The values of the different parameters were implemented in the HERWIG Monte Carlo. To summarize, we input into HERWIG the NLL-BFKL parton-level cross section which depends on $p_T$ and $\Delta\eta$, and the output depends on $\Delta\eta_g$, $\Delta\eta_J$, and the jets transverse momenta $p_{T_1}$ and $p_{T_2}$. Further integrations of these kinematic variables are performed to obtain the different observables discussed in the next section, taking into account experimental cuts.

\section{Comparaison with Tevatron data}

\begin{figure}[t]
\begin{center}
\epsfig{file=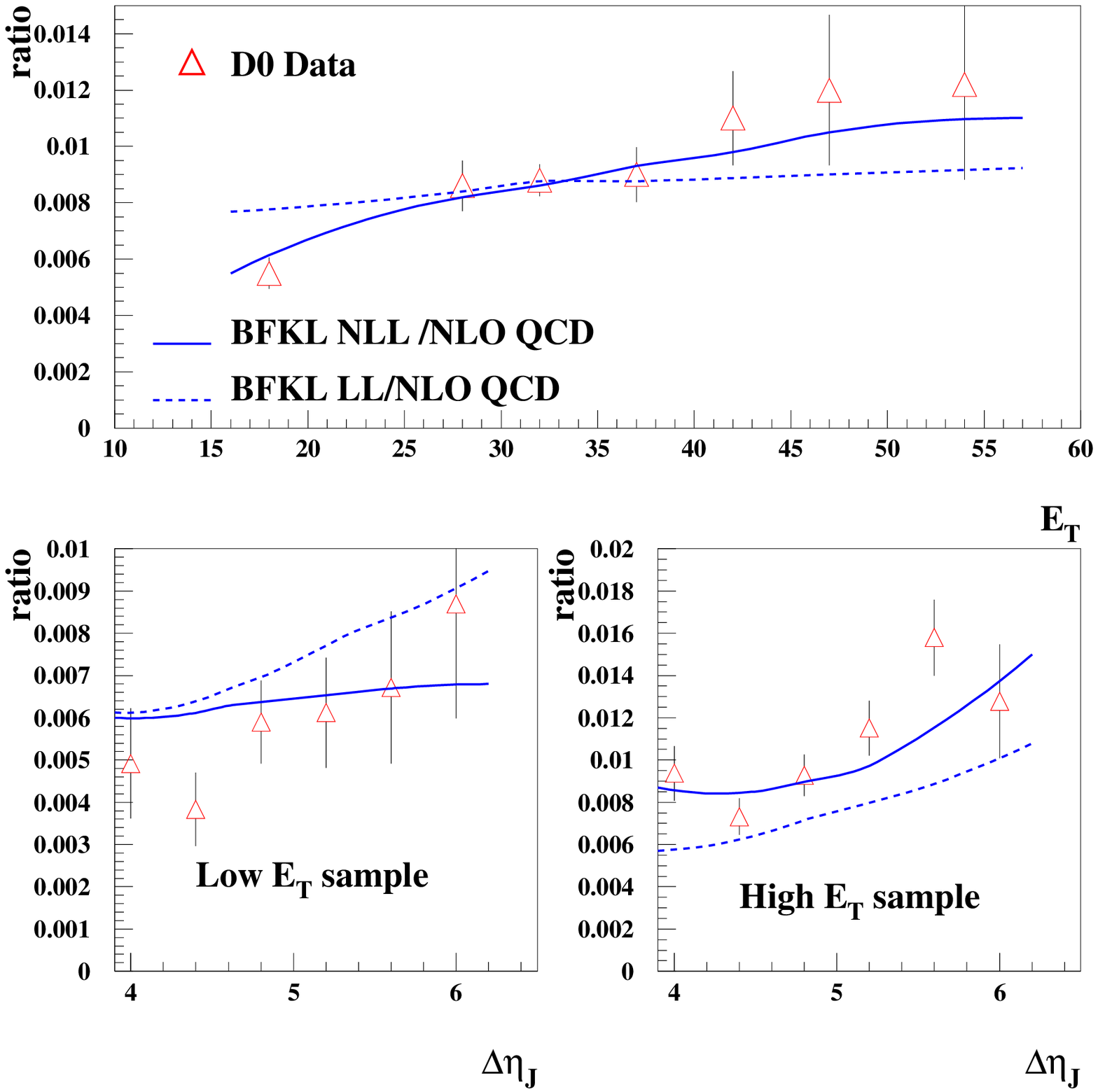,width=8.9cm}
\hfill
\epsfig{file=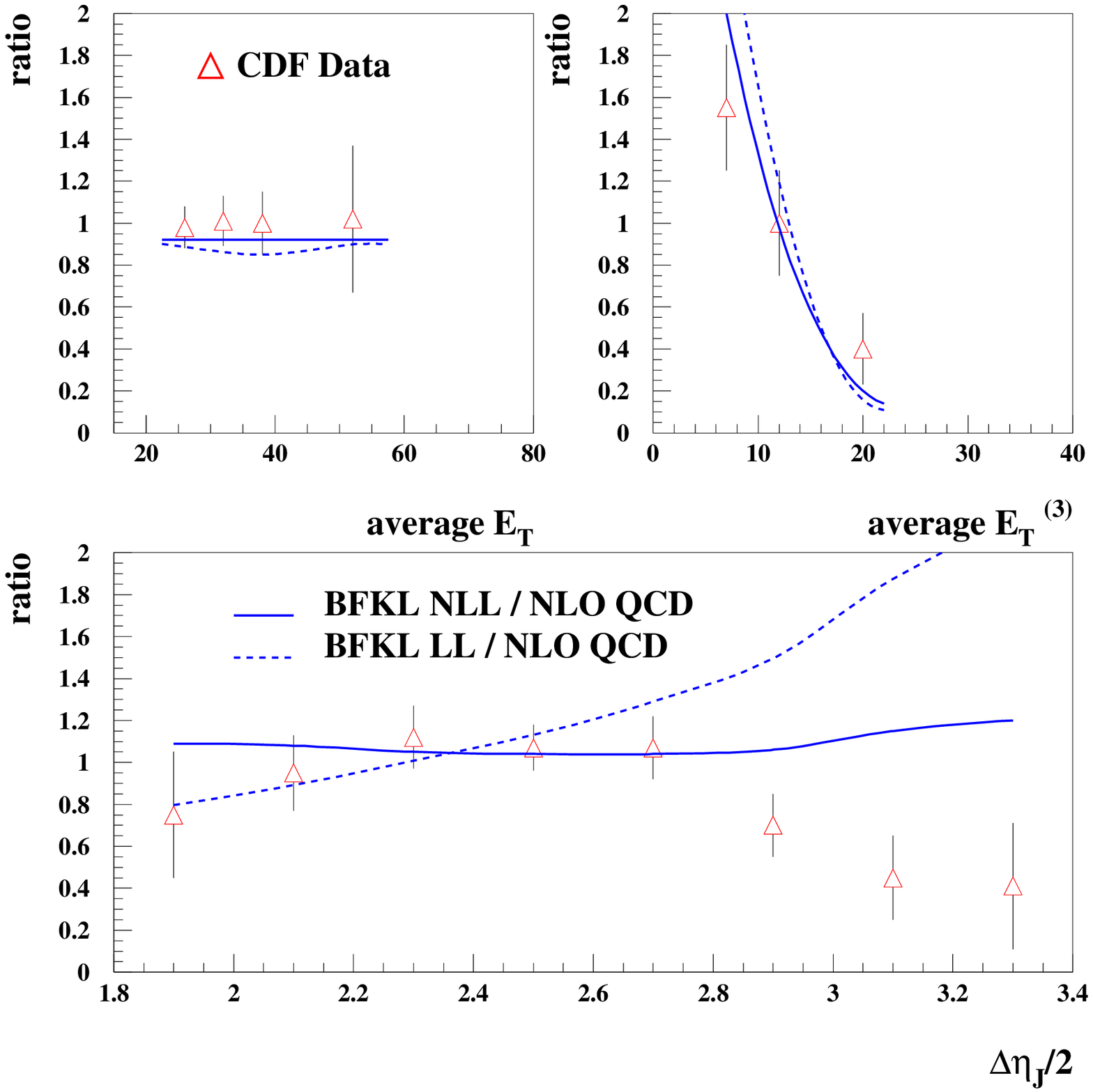,width=8.9cm}
\end{center}
\caption{Comparisons between the D0 (left) and CDF (right) measurements of the jet-gap-jet event ratio with the NLL- and LL-BFKL calculations. The NLL calculation is in fair agreement with the data while the LL one leads to a worse description.}
\label{tevatron}
\end{figure}

The D0 collaboration has performed a measurement of the jet-gap-jet event ratio, defined as the ratio of the jet-gap-jet cross section to the inclusive di-jet cross section, as a function of the transverse energy of the second-leading jet, that we denote $E_T$, and also as a function of the rapidity difference
$\Delta\eta_J$ between the two leading jets \cite{d0}. At least two jets are reconstructed in the D0 calorimeter with $E_T> 15\ \mbox{GeV}$ for the second leading jet. In addition, the two jets are required to be in the forward regions and in opposite hemispheres, by requesting $1.9<|\eta_{1,2}|<4.1$ and
$\eta_1 \eta_2 <0$. The difference in rapidity between both jets $\Delta\eta_J$ was imposed to be larger than 4, and a rapidity gap between at least $\eta=-1$ and $\eta=1$ is required. The data are presented as a function of the second-leading-jet $E_T$, or as a function of $\Delta\eta_J$ in which case a low-$E_T$ and a high-$E_T$ jet samples were used (low $E_T$ means $15<E_T<25\ \mbox{GeV}$ and high $E_T$ means $E_T>30\ \mbox{GeV},$ those cuts applying to both jets).

To compare directly the D0 measurement with the NLL-BFKL calculation implemented in HERWIG, we compute the following ratio
\be
R=\frac{NLL~BFKL~Herwig}{Jet~Herwig}\times\frac{LO~QCD}{NLO~QCD}\ ,
\ee
where $NLL~BFKL~Herwig$ and $Dijet~Herwig$ are the jet-gap-jet and the inclusive di-jet cross sections obtained with HERWIG, respectively. To take into account NLO QCD effects, we also correct the ratio
$R$ by the LO/NLO QCD di-jet cross section ratio obtained with the NLOJet++ program \cite{nlojet}. The same method applies for the LL-BFKL cross section calculation. The comparison between our calculations and the D0 data is given in the left plot of Fig.~\ref{tevatron}, after the overall normalization was adjusted in the two cases (LL and NLL). We find that there is a good agreement between the NLL-BFKL calculation and the data whereas the LL-BFKL calculation leads to an $E_T$ dependence which is too flat. Moreover, in the LL case it is difficult to accomodate all the data with a single overall normalization factor, while in the NLL case this is not a problem.

The comparison with CDF data is given in the right plot of Fig.\ref{tevatron}, in this case the normalization of the different data sets is arbitrary. The CDF collaboration also measured the jet-gap-jet cross section requesting a gap between -1 and 1 in rapidity, but used a higher jet $E_T$ threshold of 20 GeV and a lower acceptance in jet rapidity between 1.8 to 3.5, compared to the D0 measurement. The CDF requirement on the minimum rapidity interval between the two jets is then $\Delta\eta_J>3.6$, compared to 4 in the case of D0.
CDF measured the jet-gap-jet event ratio as a function of the average jet transverse momentum (this is what $E_T$ denotes in the case of CDF), the average transverse momentum of the third jet $E_T^{(3)}$ when there is one in the event, and also as a function of $\Delta\eta_J$. The conclusion remains the same as in the case of D0 data, namely that the NLL-BFKL formalism leads to a better description than the LL one. However, it is worth noticing that we are not able to describe the full $\Delta\eta_J$ dependence and especially the decrease at high $\Delta\eta_J$, which is somehow in disagreement with the D0 measurement. Further measurements in progress in the CDF collaboration will be very useful to understand these differences. Parton showering and hadronization effects are crucial in order to obtain this level of agreement with data.

\begin{figure}
\begin{center}
\epsfig{file=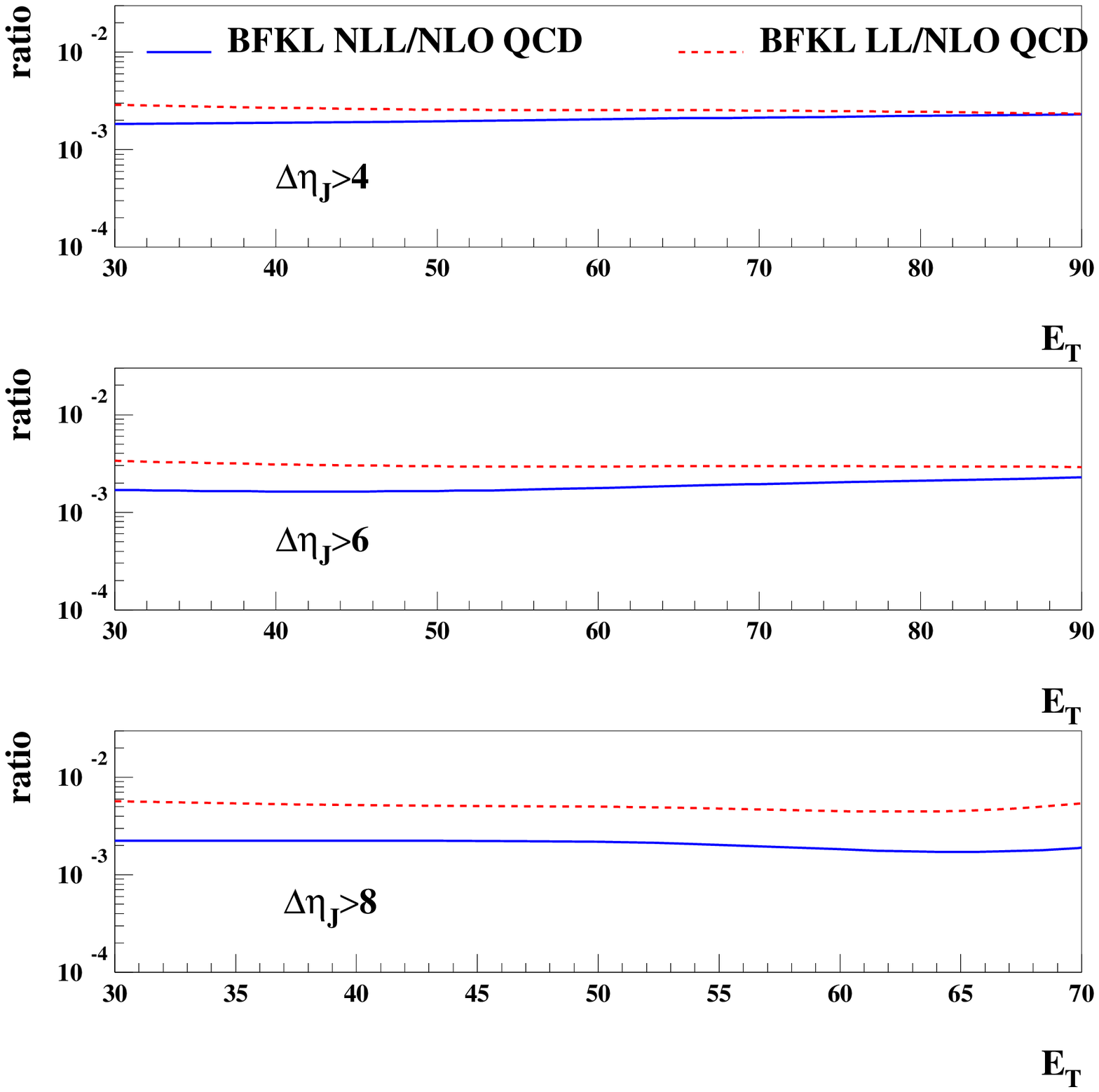,width=8.7cm}
\hfill
\epsfig{file=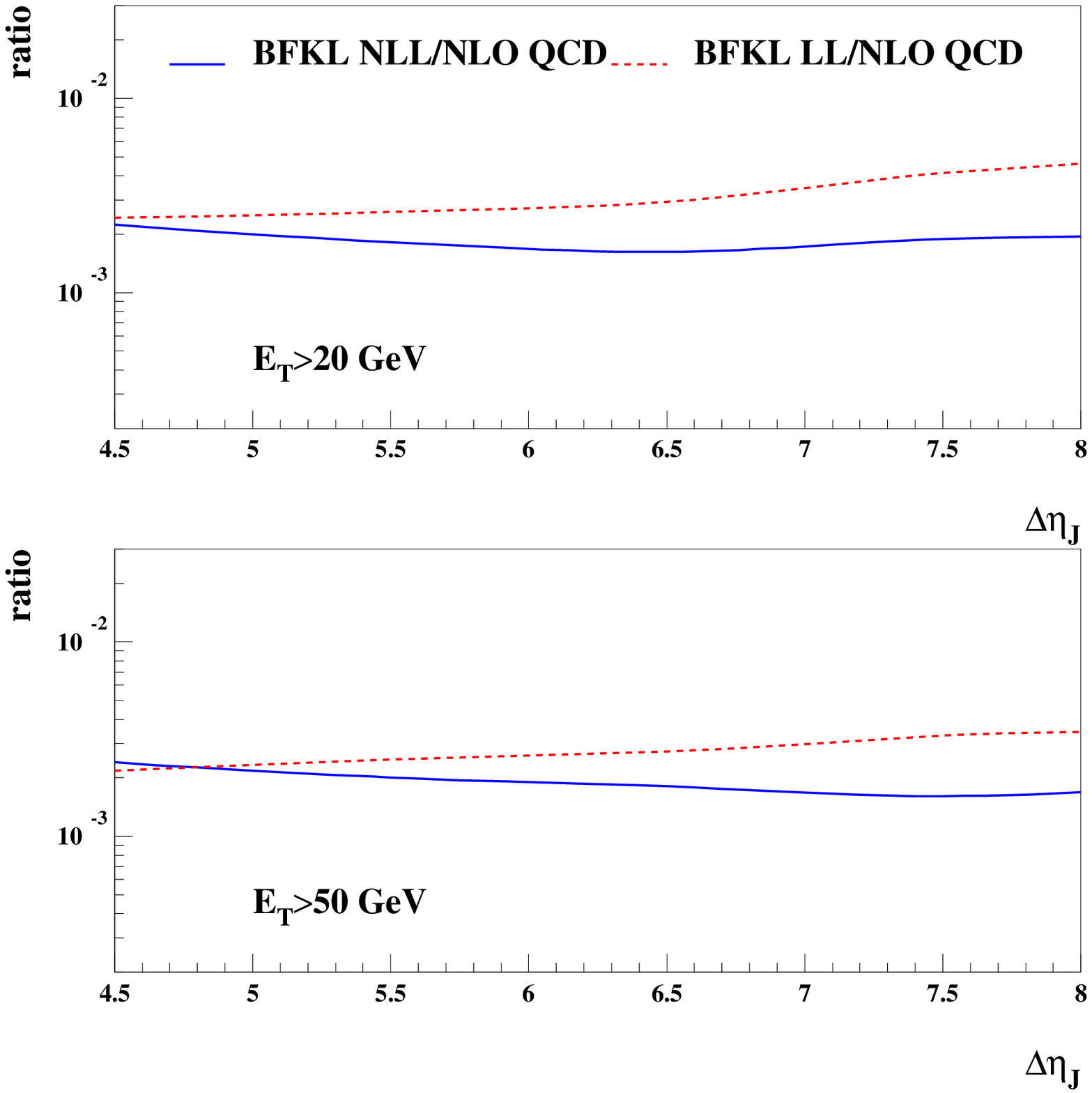,width=8.7cm}
\end{center}
\caption{Predictions of our model for the ratio of the jet-gap-jet to the inclusive-jet cross section at the LHC, as a function of the second-leading-jet transverse energy $E_T$ (left), and of the rapidity difference between the two leading jets $\Delta\eta_J$ (right).}
\label{lhc}
\end{figure}

\section{Predictions for the LHC}

Using the normalizations obtained from the fits to the D0 data, we are able to predict the
jet-gap-jet event ratio at the LHC. In doing so, we also take into account the fact that the gap-survival probability is smaller by a factor 10/3 (this is estimated for a collision energy of $\sqrt{s}=14$ TeV).
Requesting both jets to have $E_T>20$ GeV, and the jet rapidities to obey $2<|\eta_{1,2}|<5$ and
$\eta_1 \eta_2 <0$, the values of jet-gap-jet event ratios are shown in Fig.~\ref{lhc}, as a function of $E_T$ for different $\Delta\eta_J$ ranges (left plot), and as a function of $\Delta\eta_J$ for different $E_T$ ranges (right plot). The main feature of the predictions of our model is that jet-gap-jet event ratio is about 0.002 and does not vary a lot with $E_T$ or $\Delta\eta_J$, except for the LL-BFKL predictions which increase a bit with $\Delta\eta_J$. By contrast, the parton-level predictions obtained in \cite{us} featured an increase of the ratio with both $E_T$ or $\Delta\eta_J$.

Let us add a word of caution about these predictions. We have assumed a specific value for the gap survival probability at the LHC (namely 0.03), however this value shows large theoretical uncertainties \cite{gapsurvival}. It will be measured by the LHC experiments and our cross section predictions will have to be modified once such measurements are performed. For instance, our cross section should be reduced by a factor 10 if the survival probability is found to be ten times smaller than 0.03 for a center-of-mass energy of 14 TeV. In addition, our prediction assumes that the inclusive jet cross section at the LHC can be correctly described by the Herwig Monte Carlo, as is the case at the Tevatron \cite{Abazov:2004hm}. This can be again tested at the LHC using the first data. One first indication for a center-of-mass energy of 7 TeV was given by the measurement of the difference in azimuthal angle in dijet events \cite{lhcprelim} which seems to be well described by the Herwig Monte Carlo. However, the same measurement as well as cross section comparison between data and MC need to be done at a higher center-of-mass energy at 14 TeV to make sure that the Herwig Monte Carlo can describe the inclusive jet measurement. If some discrepancy need to be accounted for, this has to be taken into account as well in our prediction of the jet gap jet cross section ratio.

\section{Conclusions}

We have embedded the parton-level NLL-BFKL calculation of \cite{us} into the HERWIG Monte Carlo program, in order to obtain hadron-level results for the jet-gap-jet cross-section in hadron-hadron collisions, corresponding to the production of two high-$p_T$ jets around a large rapidity gap. The NLL-BFKL effects are implemented through a renormalization-group improved kernel in the S4 scheme, while the Mueller-Tang prescription is used to couple the BFKL Pomeron to colored partons, described with only LO impact factors.

After adjusting one parameter, the overall normalization, the NLL-BFKL calculation is able to describe all Tevatron data, except the higher end of the $\Delta\eta_J$ dependence measured by CDF. This still provides an improvement compared to the LL-BFKL calculation (obtained with the fixed value of the coupling
$\bar\alpha=0.16$), which in addition features an $E_T$ dependence that is too flat compared to the D0 data. We presented predictions which could be tested at the LHC, for the same jet-gap-jet event ratio measured at 
the Tevatron, but for larger rapidity gaps.

We noticed that going from parton-level to hadron-level is necessary in order to obtain a global description of the Tevatron data, hence our results supersede those of Ref.~\cite{us}. Considering LHC predictions,
hadron-level calculations show almost no dependence of the jet-gap-jet event ratio (which is about 0.002 in the kinematic range we considered, and for the experimental cuts we used), while parton-level calculations showed an increase of this ratio with both $E_T$ and $\Delta\eta_J$. This should provide a strong test of the BFKL regime.

\end{document}